\documentclass[11pt,english]{article}
\usepackage[T1]{fontenc}
\usepackage[latin9]{inputenc}
\usepackage{color}
\usepackage{amsmath}
\usepackage{amssymb}
\usepackage{graphicx}
\usepackage{setspace}
\usepackage[authoryear]{natbib}
\makeatletter

\providecommand{\tabularnewline}{\\}

\date{}
\usepackage{tikz}
\newtheorem{theorem}{Theorem}
\newtheorem{remark}{Remark}

\newtheorem{condition}{Condition}

\newtheorem{assumption}{Assumption}

\newcommand*{\indep}{%
  \mathbin{%
    \mathpalette{\@indep}{}%
  }%
}
\newcommand*{\nindep}{%
  \mathbin{
    \mathpalette{\@indep}{\not}
  }%
}
\newcommand*{\@indep}[2]{%
  \sbox0{$#1\perp\m@th$}
  \sbox2{$#1=$}
  \sbox4{$#1\vcenter{}$}
  \rlap{\copy0}
  \dimen@=\dimexpr\ht2-\ht4-.2pt\relax
  \kern\dimen@
  {#2}%
  \kern\dimen@
  \copy0 
} 

\newcommand*{\pr}{\mathrm{pr}}
\newcommand*{\var}{\mathrm{var}}

\newcommand*{\iptw}{\mathrm{IPTW}}
\newcommand*{\HT}{\mathrm{HT}}
\newcommand{\T}{\mathrm{\scriptscriptstyle T}}
\newcommand*{\plim}{\mathrm{plim}}
\newcommand*{\fe}{\mathrm{fix}}
\newcommand*{\re}{\mathrm{ran}}
\newcommand*{\N}{\mathcal{N}}

\makeatother

\usepackage{babel}
\begin{document}

\title{Propensity score weighting for causal inference with clustered data }

\author{Shu Yang\thanks{Department of Statistics, North Carolina State University, Raleigh,
North Carolina 27695, U.S.A. syang24@ncsu.edu} }
\maketitle
\begin{abstract}
Propensity score weighting is a tool for causal inference to adjust
for measured confounders in observational studies. In practice, data
often present complex structures, such as clustering, which make propensity
score modeling and estimation challenging. In addition, for clustered
data, there may be unmeasured cluster-specific variables that are
related to both the treatment assignment and the outcome. When such
unmeasured cluster-specific confounders exist and are omitted in the
propensity score model, the subsequent propensity score adjustment
may be biased. In this article, we propose a calibration technique
for propensity score estimation under the latent ignorable treatment
assignment mechanism, i.e., the treatment-outcome relationship is
unconfounded given the observed covariates and the latent cluster
effects. We then provide a consistent propensity score weighting estimator
of the average treatment effect when the propensity score and outcome
follow generalized linear mixed effects models. The proposed propensity
score weighting estimator is attractive, because it does not require
specification of functional forms of the propensity score and outcome
models, and therefore is robust to model misspecification. The proposed
weighting method can be combined with sampling weights for an integrated
solution to handle confounding and sampling designs for causal inference
with clustered survey data. In simulation studies, we show that the
proposed estimator is superior to other competitors. We estimate the
effect of School Body Mass Index Screening on prevalence of overweight
and obesity for elementary schools in Pennsylvania. 

\end{abstract}

\section{Introduction}

Observational studies are often used to infer causal effects in medical
and social science studies. In observational studies, there often
is confounding by indication: some covariates are predictors of both
the treatment and outcome. One implication is that the covariate distributions
differ between the treatment groups. Under the assumption of ignorable
treatment assignment and that all confounders are observed, the causal
effect of treatments can be obtained by comparing the outcomes for
units from different treatment groups, adjusting for the observed
confounders. \citet{rosenbaum1983central} further claimed the central
role of the propensity score, and showed that adjusting for the propensity
score removes confounding bias. An extensive literature thereafter
proposed a number of estimators based on the propensity score. In
particular, propensity score weighting can be used to create a weighted
population where the covariate distributions are balanced between
the treatment groups, on average. Therefore, the comparison between
the weighted outcomes has a causal interpretation; see \citet{imbens2015causal}
for a textbook discussion. 

Propensity score weighting has been mainly developed and applied in
settings with independently and identically distributed (i.i.d.) data.
However, in many research area, data often present complex structures,
such as clustering. For a motivating example, we examine the 2007\textendash 2010
body mass index (BMI) surveillance data from Pennsylvania Department
of Health to estimate the effect of School Body Mass Index Screening
(SBMIS) on the annual overweight and obesity prevalence in elementary
schools in Pennsylvania. The data set includes $493$ school districts
in Pennsylvania, which are clustered by two factors: location (rural,
suburban, and urban), and population density (low, median, and high).
In this data set, $63\%$ of schools implemented SBMIS, and the percentages
of schools implemented SBMIS across the clusters are from $45\%$
to $70\%$, indicating cluster-level heterogeneity of treatment. Moreover,
even if we collect a rich set of unit-level covariates, there may
be unobserved cluster effects that are related to both the treatment
and outcome. In our motivating example, we have school-level covariates
including the baseline prevalence of overweight and obesity and percentage
of reduced and free lunch. However, certain key health factors, such
as accessibility to and quality of care, socioeconomic   and environmental
variables, which can be very different across clusters, are preceivably
important factors for children's obesity rate and implementing prevention
screening strategy. Unfortunately, these cluster-specific confounders
are not available. When such unmeasured confounders exist and are
omitted in the propensity score model, the subsequent analysis may
be biased. Although important for empirical practice, much less work
has been done for causal inference with clustered data subject to
unmeasured cluster-specific confounding. 

The goal of this article is to develop a novel propensity score weighting
method for causal inference with clustered data in the presence of
unmeasured cluster-level confounders. Natural models for the propensity
score and outcome are generalized linear mixed effects models, where
cluster-level confounding is captured via cluster random effect terms.
Prior to our work, \citet{li2013propensity} investigated the performance
of the propensity score weighting estimators under generalized linear
fixed/mixed effects models for the propensity score and outcome. However,
their approach requires correct specification of functional forms
of the propensity score and outcome models. In this article, we provide
a robust construction of inverse propensity score weights under the
latent ignorable treatment assignment mechanism, i.e., the treatment-outcome
relationship is unconfounded given the observed covariates and the
latent cluster effects. The key insight is based on the central role
of the propensity score in balancing the covariate distributions between
the treatment groups. For propensity score estimation, we then adopt
the calibration technique and impose balancing constraints for moments
of the observed and latent cluster-specific confounders between the
treatment groups. These constraints eliminate confounding biases.
Under certain regularity conditions, the propensity score weighting
estimator of the average treatment effect is consistent when the propensity
score and outcome follow generalized linear mixed effects models.
The proposed propensity score weighting estimator does not require
correct specification of functional forms of the treatment and outcome
models, and therefore is robust to model misspecification. 

\section{Basic Setup\label{sec:Basic-Setup}}

\subsection{Observed data structure }

To fix the ideas, we first focus on clustered data. The extension
to clustered survey data will be addressed in Section \ref{sec:Extension-to-survey}.
Suppose that the sample consists of $m$ clusters, and cluster $i$
includes $n_{i}$ units. Denote the sample size by $n=\sum_{i=1}^{m}n_{i}$.
For unit $j$ in cluster $i$, we observe a $p$-dimensional vector
of pre-treatment variables $X_{ij}$, which may include the observed
individual and cluster characteristics, a binary treatment variable
$A_{ij}\in\{0,1\}$, with $0$ and $1$ being the labels of control
and active treatments, respectively, and lastly an outcome variable
$Y_{ij}$. 

\subsection{Models and assumptions \label{subsec:Potential-outcomes}}

We use the potential outcomes framework \citep{rubin1974estimating}.
Assume that each unit has two potential outcomes: $Y_{ij}(0)$, the
outcome that would be realized, possibly contrary to the fact, had
the unit received the control treatment, and $Y_{ij}(1)$, the outcome
that would be realized, possibly contrary to the fact, had the  unit
received the active treatment. This notation implicitly assumes that
there is no interference between units and no versions of each treatment
(\citealp{rubin1978bayesian}). The observed outcome is $Y_{ij}=Y_{ij}(A_{ij})$
(\citealp{rubin1974estimating}). Suppose that $\{A_{ij},X_{ij},Y_{ij}(0),Y_{ij}(1):i=1,\ldots,m;j=1,\ldots n_{i}\}$
follows an infinite super-population distribution. Our goal is to
estimate the average treatment effect, $\tau=E\{Y_{ij}(1)-Y_{ij}(0)\}$,
where the expectation is taken with respect to the super-population
distribution. In the binary case, $\tau$ is called the causal risk
difference.

The fundamental problem is that not all potential outcomes can be
observed for each unit in the sample; only one potential outcome,
the outcome corresponding to the treatment the unit actually followed,
can be observed. With unstructured i.i.d. data, \citet{rubin1974estimating}
described the following assumption for identifying average treatment
effect.

\begin{assumption}[Ignorability]\label{asump:strongIg} $\{Y_{ij}(0),Y_{ij}(1)\}\indep A_{ij}\mid X_{ij}.$

\end{assumption} 

Assumption \ref{asump:strongIg} indicates that all confounders are
observed. For clustered data, confounding may vary across clusters
and related to some cluster-specific variables that are not always
observable. In these cases, Assumption \ref{asump:strongIg} does
not hold. We assume a cluster-specific latent effect $U_{i}$ that
summarizes the effect of unobserved cluster-level confounders, and
consider the following modified ignorability assumption.

\begin{assumption}[Latent ignorability]\label{asump:LatentIgnorability}
$\{Y_{ij}(0),Y_{ij}(1)\}\indep A_{ij}\mid X_{ij},U_{i}.$

\end{assumption} 

Under Assumption \ref{asump:LatentIgnorability}, $\pr\{A_{ij}=1\mid X_{ij},U_{i},Y_{ij}(0),Y_{ij}(1)\}=\pr(A_{ij}=1\mid X_{ij},U_{i}),$
which is the propensity score. We assume the numbers of units received
the active and control treatments are nonzero in each cluster; otherwise,
there exist units in some clusters for which we can not estimate the
average treatment effect without extrapolation assumptions. 

\begin{assumption}[Positivity]\label{asump:positivity}There exist
constants $\underline{e}$ and $\bar{e}$ such that, with probability
$1$, $0<\underline{e}<\pr(A_{ij}=1\mid X_{ij},U_{i})<\bar{e}<1$. 

\end{assumption} 

Under Assumption \ref{asump:LatentIgnorability}, we can write the
conditional distribution of $\{(A_{ij},Y_{ij}):i=1,\ldots,m;j=1,\ldots n_{i}\}$
given $\{(X_{ij},U_{i}):i=1,\ldots,m;j=1,\ldots n_{i}\}$ as 
\begin{multline*}
\prod_{i=1}^{m}\prod_{j=1}^{n_{i}}\left\{ f_{1}(Y_{ij}\mid X_{ij},U_{i})\pr(A_{ij}=1\mid X_{ij},U_{i})\right\} ^{A_{ij}}\\
\times\left[f_{0}(Y_{ij}\mid X_{ij},U_{i})\{1-\pr(A_{ij}=1\mid X_{ij},U_{i})\}\right]^{1-A_{ij}},
\end{multline*}
where $f_{a}(\cdot\mid X_{ij},U_{i})$ is a conditional distribution
of $Y_{ij}(a)$ given $(X_{ij},U_{i})$, for $a=0,1$. 

There are two different model specifications regarding the cluster-level
effects. The fixed effects model treats $U_{i}$ as fixed but unknown
parameters across clusters. In this fixed-effects approach, the treatment
assignment mechanism is an ignorable process, which complies with
Assumption \ref{asump:strongIg} given that $X_{ij}$ stacks all observed
confounders and cluster-specific dummy variables. On the other hand,
the random effects model treats $U_{i}$ as random and i.i.d. drawn
from a distribution. The difference between the two modeling strategies
has been addressed in both statistics and econometrics literature;
see, e.g., \citet{baltagi1995econometric} and \citet{wooldridge2010econometric}.
Briefly, there are two main considerations: one on statistical consideration
and the other on logic consideration. First, if the number of clusters
is relatively large, the estimator from the fixed effects approach
becomes inconsistent in propensity score estimation \citep{wallace1969use}.
In this case, the random effects approach is preferred. Second, the
fixed effects approach does not make distributional assumptions of
the cluster-specific effects; whereas, the random effects approach
assumes that $U_{i}$ is random and i.i.d. drawn from a distribution.
To justify this random effects assumption, we can use the exchangeability
criterion of \citet{chamberlain1984panel}. If the $U_{i}$'s can
be randomly permuted to ensure exchangeability; a version of de Finetti's
theorem implies then that they are i.i.d. as draws from an appropriately
defined distribution. In the case of evaluating the causal effect
of SBMIS on children's obesity, this assumption implies that unobserved
cluster characteristics that influence both implementing SBMIS and
children's obesity are not correlated with school characteristics
that are included in the models. 

\begin{figure}
\begin{centering}
\includegraphics[scale=0.4]{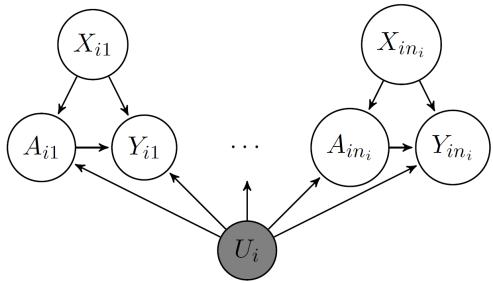}
\par\end{centering}
\caption{\label{fig:1}A direct acyclic graph (DAG) illustrating the dependence
of variables under our assumptions for cluster $i$. $A$ is the treatment,
$Y$ is the outcome, $X$ is the measured confounders, and $U$ is
the unmeasured cluster-level confounders.}
\end{figure}

In this article, we assume that the $U_{i}$'s are random variables
and independently follow a certain distribution. Figure \ref{fig:1}
provides a directed acyclic graph \citep{pearl2009causality} that
implies the dependence of variables under our assumptions in cluster
$i$. 
 We now posit generalized linear mixed effects models for $f_{a}(Y_{ij}\mid X_{ij},U_{i})$
and $\pr(A_{ij}=1\mid X_{ij},U_{i})$. To be specific, we suppose
that $Y_{ij}(a)$ follows a generalized linear mixed effects model
with a random effect $U_{i}$ as

\begin{equation}
g\{\mu_{ij}(a)\}=X_{ij}^{\T}\beta_{a}+U_{i},\label{eq:glmm}
\end{equation}
where $\mu_{ij}(a)=E\{Y_{ij}(a)\mid X_{ij},U_{i}\}$, $g(\cdot)$
is the link function, and $\beta_{a}$ is a $p$-dimensional vector
of fixed effects of $X_{ij}$. Similarly, we assume that $A_{ij}$
given $(X_{ij},U_{i})$ follows a generalized linear mixed effects
model as 
\begin{equation}
\pr(A_{ij}=1\mid X_{ij},U_{i})=h(X_{ij}^{\T}\gamma+U_{i})\equiv e(X_{ij},U_{i};\eta),\label{eq:glmm_ps}
\end{equation}
where $h(\cdot)$ is the inverse link function, and $\gamma$ is a
$q$-dimensional vector of parameters. 

\subsection{Inverse probability of treatment weighting estimator}

To estimate the average treatment effect $\tau$, let $\nu=(U_{1},\ldots,U_{m})$
denote the vector of random effects. The inverse propensity score
or probability of treatment weighting (IPTW) estimator of $\tau$
can be expressed as
\begin{equation}
\hat{\tau}_{\iptw}(\nu,\eta)=\frac{1}{n}\sum_{i=1}^{m}\sum_{j=1}^{n_{i}}\left\{ \frac{A_{ij}Y_{ij}}{e(X_{ij},U_{i};\eta)}-\frac{(1-A_{ij})Y_{ij}}{1-e(X_{ij},U_{i};\eta)}\right\} .\label{eq:iptw}
\end{equation}
Under Assumptions \ref{asump:LatentIgnorability} and \ref{asump:positivity},
if the propensity score is known, it is straightforward to verify
that $\hat{\tau}_{\iptw}(\nu,\eta)$ is unbiased for $\tau$. Moreover,
if it is unknown but depends only on fixed parameters, $\hat{\tau}_{\iptw}(\nu,\eta)$
with the consistently estimated propensity score is asymptotically
unbiased for $\tau$. The challenge with clustered data is that $\hat{\tau}_{\iptw}(\nu,\eta)$
depends on a growing number of unobserved random effects with the
number of clusters. To resolve this issue, there are several options:
\begin{description}
\item [{(i)}] Weight based on predicted random effects; i.e., based on
the generalized linear mixed effects model in (\ref{eq:glmm_ps}),
predict the propensity score as $e(X_{ij},\hat{U}_{i}^{\re};\hat{\eta})$,
where $\hat{U}_{i}^{\re}$ is the mode of a predictive distribution
for $U_{i}$ given the observed $A_{ij}$ and $X_{ij}$, and $\hat{\eta}$
is the maximum likelihood estimator of $\eta$. 
\item [{(ii)}] Weight based on estimated fixed effects; i.e., treat the
$U_{i}$'s in model (\ref{eq:glmm_ps}) as fixed effects, and estimate
the propensity score as $e(X_{ij},\hat{U}_{i}^{\fe};\hat{\eta})$,
where $\hat{U}_{i}^{\fe}$ and $\hat{\eta}$ are maximum likelihood
estimators. 
\end{description}
Let $\hat{\tau}_{\iptw}(\nu,\eta)$ in (\ref{eq:iptw}) be denoted
as $\hat{\tau}_{\re}$ or $\hat{\tau}_{\fe}$ when the propensity
score is predicted under option (i) or estimated under option (ii),
respectively. The two approaches suffer several drawbacks. Firstly,
to obtain $\hat{\tau}_{\re}$ often involves numerical integration,
and therefore can be computationally heavy. Secondly, the predicted
value of the propensity score does not guarantee the balance of covariate
distributions between the treatment groups, due to the shrinkage of
random effects toward zero. Lastly, $\hat{\tau}_{\fe}$ does not yield
a consistent estimator for $\tau$ for small $n_{i}$ \citep{skinner2011inverse}.

\section{Proposed methodology }

\subsection{The new IPTW estimator }

To motivate our estimation of the propensity score, we note 
\begin{equation}
E\left\{ \frac{A}{e(X,U)}\left(\begin{array}{c}
X\\
U
\end{array}\right)\right\} =E\left[E\left\{ \frac{A}{e(X,U)}\mid X,U\right\} \left(\begin{array}{c}
X\\
U
\end{array}\right)\right]=E\left\{ \left(\begin{array}{c}
X\\
U
\end{array}\right)\right\} ,\label{eq:p1}
\end{equation}
and 
\begin{equation}
E\left\{ \frac{1-A}{1-e(X,U)}\left(\begin{array}{c}
X\\
U
\end{array}\right)\right\} =E\left[E\left\{ \frac{1-A}{1-e(X,U)}\mid X,U\right\} \left(\begin{array}{c}
X\\
U
\end{array}\right)\right]=E\left\{ \left(\begin{array}{c}
X\\
U
\end{array}\right)\right\} .\label{eq:p2}
\end{equation}
(\ref{eq:p1}) and (\ref{eq:p2}) clarify the central role of the
propensity score in balancing the covariate distributions between
the treatment groups in the super-population. For simplicity of presentation,
let $e_{ij}$ be the propensity score for unit $j$ in cluster $i$.
We consider the propensity score estimate $\hat{e}_{ij}$ that satisfies
the empirical version of (\ref{eq:p1}) and (\ref{eq:p2}):
\begin{equation}
\sum_{i=1}^{m}\sum_{j=1}^{n_{i}}\frac{A_{ij}}{\hat{e}_{ij}}X_{ij}=\sum_{i=1}^{m}\sum_{j=1}^{n_{i}}\frac{1-A_{ij}}{1-\hat{e}_{ij}}X_{ij}=\sum_{i=1}^{m}\sum_{j=1}^{n_{i}}X_{ij},\label{eq:c1}
\end{equation}

\begin{equation}
\sum_{j=1}^{n_{i}}\frac{A_{ij}}{\hat{e}_{ij}}=\sum_{j=1}^{n_{i}}\frac{1-A_{ij}}{1-\hat{e}_{ij}}=\sum_{j=1}^{n_{i}}1=n_{i},\quad(i=1,\ldots,m).\label{eq:c2}
\end{equation}
To obtain the propensity score estimate that achieves (\ref{eq:c1})
and (\ref{eq:c2}), we use the calibration technique in the following
steps:
\begin{description}
\item [{Step$\ $1.}] \label{step1} Obtain an initial propensity score
estimate $\hat{e}_{ij}^{0}$ under some working propensity score model,
e.g. a logistic linear model fitted to $(A_{ij},X_{ij})$. This in
turn provides an initial set of inverse propensity score weights,
$\mathbb{W}^{0}=\{d_{ij};i=1,\ldots,m,j=1,\ldots,n_{i}\}$, where
$d_{ij}=1/e_{ij}^{0}$ if $A_{ij}=1$ and $d_{ij}=1/(1-e_{ij}^{0})$
if $A_{ij}=0$.
\item [{Step$\ $2.}] \label{step2} Modify the initial set of weights
$\mathbb{W}^{0}$ to a new set of weights $\mathbb{W}=\{\alpha_{ij};i=1,\ldots,m,j=1,\ldots,n_{i}\}$
by minimizing the Kullback-Leibler distance \citep{kullback1951information}
of $\mathbb{W}^{0}$ and $\mathbb{W}$:
\begin{equation}
\sum_{i=1}^{m}\sum_{j=1}^{n_{i}}G(\alpha_{ij},d_{ij})=\sum_{i=1}^{m}\sum_{j=1}^{n_{i}}\alpha_{ij}\log\frac{\alpha_{ij}}{d_{ij}},\label{eq:KL}
\end{equation}
subject to (\ref{eq:c1}) and (\ref{eq:c2}). By Lagrange Multiplier,
the minimizer of (\ref{eq:KL}) subject to (\ref{eq:c1}) and (\ref{eq:c2})
is
\begin{multline}
\alpha_{ij}(\lambda_{1},\lambda_{2})=\frac{n_{i}A_{ij}d_{ij}\exp(\lambda_{1}X_{ij}A_{ij})}{\sum_{j=1}^{n_{i}}A_{ij}d_{ij}\exp(\lambda_{1}X_{ij}A_{ij})}\\
+\frac{n_{i}(1-A_{ij})d_{ij}\exp\{\lambda_{2}X_{ij}(1-A_{ij})\}}{\sum_{j=1}^{n_{i}}(1-A_{ij})d_{ij}\exp\{\lambda_{2}X_{ij}(1-A_{ij})\}},\label{eq:alpha_ij}
\end{multline}
where $(\lambda_{1},\lambda_{2})^{\T}$ is the solution to the following
equation
\begin{multline}
\hat{Q}(\lambda_{1},\lambda_{2})=\left(\begin{array}{c}
\hat{Q}_{1}(\lambda_{1},\lambda_{2})\\
\hat{Q}_{2}(\lambda_{1},\lambda_{2})
\end{array}\right)\\
=\left(\begin{array}{c}
n^{-1}\sum_{i=1}^{m}\sum_{j=1}^{n_{i}}\left\{ A_{ij}\alpha_{ij}(\lambda_{1},\lambda_{2})-1\right\} X_{ij}\\
n^{-1}\sum_{i=1}^{m}\sum_{j=1}^{n_{i}}\left\{ (1-A_{ij})\alpha_{ij}(\lambda_{1},\lambda_{2})-1\right\} X_{ij}
\end{array}\right)=0.\label{eq:hat Q}
\end{multline}
\item [{Step$\ $3.}] Obtain the propensity score estimate as $\hat{e}_{ij}=\alpha_{ij}(\hat{\lambda}_{1},\hat{\lambda}_{2})^{-A_{ij}}\{1-\alpha_{ij}(\hat{\lambda}_{1},\hat{\lambda}_{2})\}^{-1+A_{ij}}$. 
\end{description}
Finally, our proposed IPTW estimator is 
\begin{equation}
\hat{\tau}_{\mathrm{\iptw}}=\frac{1}{n}\sum_{i=1}^{m}\sum_{j=1}^{n_{i}}\left\{ \frac{A_{ij}Y_{ij}}{\hat{e}_{ij}}-\frac{(1-A_{ij})Y_{ij}}{1-\hat{e}_{ij}}\right\} .\label{eq:proposed estimator}
\end{equation}

\begin{remark} 

Calibration has been used in many scenarios\textcolor{black}{.} \textcolor{black}{In
survey sampling, calibration is widely used to integrate auxiliary
information in estimation or handle nonresponse; see, e.g., \citet{wu2001model},
\citet{chen2002using}, \citet{kott2006using}, \citet{chang2008using}
and \citet{kim2016calibrated}. In causal inference, calibration has
been used such as in Constrained Empirical Likelihood \citep{qin2007empirical},
Entropy Balancing \citep{hainmueller2012entropy}, Inverse Probability
Tilting \citep{graham2012inverse}, and Covariate Balance Propensity
Score of \citet{imai2014covariate}. \citet{chan2015globally} showed
that estimation of average treatment effects by empirical balancing
calibration weighting can achieve global efficiency. However, all
these works were developed in settings with i.i.d. variables and they
assumed that there are no unmeasured confounders. }Our article is
the first to use calibration for causal inference with unmeasured
cluster-specific confounders.

\end{remark} 

\begin{remark} 

In Step 2 of the calibration algorithm, different distance functions,
other than the Kullback-Leibler distance, can be considered. For example,
if we choose $G(\alpha_{ij},d_{ij})=d_{ij}(\alpha_{ij}/d_{ij}-1)^{2}$,
then the minimum distance estimation leads to generalized regression
estimation \citep{park2012generalized} of the $\alpha_{ij}$'s. If
we choose $G(\alpha_{ij},d_{ij})=-d_{ij}\log(\alpha_{ij}/d_{ij})$,
then it leads to empirical likelihood estimation \citep{newey2004higher}.
We use the Kullback\textendash Leibler distance function, which leads
to exponential tilting estimation (\citealp{kitamura1997information},
\citealp{imbens1998information}, \citealp{schennach2007point}).
The advantage of using the Kullback-Leibler distance is that the resulting
weights are always non-negative. Also, with Kullback-Leibler distance,
the calibration constraint (\ref{eq:c2}) can be built into a closed
form expression for the weights, and thus avoiding solving a large
number of equations. This reduces the computation burden greatly,
when there is a large number of clusters. 

\end{remark} 

\section{Main results}

To discuss the asymptotic properties of the proposed estimator, we
assume that the number of clusters increases with $n$, i.e., $m\rightarrow\infty$,
as $n\rightarrow\infty$ , and that the cluster sample sizes satisfy
the condition that $\sup_{1\leq i\leq m}n_{i}=O(n^{1/2})$. Under
this asymptotic framework, the number of clusters increases but some
of the cluster sample sizes may remain small. We also impose certain
regularity conditions specified in the Appendix. Denote $A\cong B$
as $A=B+o_{p}(1)$, where the reference distribution is the super-population
model. 

To show the consistency of $\hat{\tau}_{\iptw}$, we first introduce
a cluster-specific mean function:
\begin{equation}
\mu_{1}(U_{i})=\frac{\int q_{1}(x,U_{i})E\{Y_{ij}(1)\mid x,U_{i}\}f(x)dx}{\int q_{1}(x,U_{i})f(x)dx},\quad q_{1}(X_{ij},U_{i})=E\left(\frac{A_{ij}}{\hat{e}_{ij}}-1\mid X_{ij},U_{i}\right),\label{eq:nu1}
\end{equation}
where $f(x)$ is the density of $X$. The key then is to note 
\begin{eqnarray}
 &  & E\left[\frac{1}{n}\sum_{i=1}^{m}\sum_{j=1}^{n_{i}}\left(\frac{A_{ij}}{\hat{e}_{ij}}-1\right)\left\{ Y_{ij}(1)-\mbox{\ensuremath{\mu}}_{1}(U_{i})\right\} \right]\nonumber \\
 & = & E\left(\frac{1}{n}\sum_{i=1}^{m}\sum_{j=1}^{n_{i}}E\left(\frac{A_{ij}}{\hat{e}_{ij}}-1\mid X_{ij},U_{i}\right)\left[E\left\{ Y_{ij}(1)\mid X_{ij},U_{i}\right\} -\mbox{\ensuremath{\mu}}_{1}(U_{i})\right]\right)\nonumber \\
 & = & E\left(\frac{1}{n}\sum_{i=1}^{m}\sum_{j=1}^{n_{i}}q_{1}(X_{ij},U_{i})\left[E\left\{ Y_{ij}(1)\mid X_{ij},U_{i}\right\} -\mbox{\ensuremath{\mu}}_{1}(U_{i})\right]\right)\rightarrow0,\label{eq:1}
\end{eqnarray}
as $m\rightarrow\infty$, which follows from the definition of $\mu_{1}(U_{i})$
in (\ref{eq:nu1}). (\ref{eq:1}) implies that 
\begin{equation}
E\left\{ \frac{1}{n}\sum_{i=1}^{m}\sum_{j=1}^{n_{i}}\left(\frac{A_{ij}}{\hat{e}_{ij}}-1\right)Y_{ij}(1)\right\} =E\left\{ \frac{1}{n}\sum_{i=1}^{m}\sum_{j=1}^{n_{i}}\left(\frac{A_{ij}}{\hat{e}_{ij}}-1\right)\mbox{\ensuremath{\mu}}_{1}(U_{i})\right\} =0,\label{eq:2}
\end{equation}
where zero follows from the constraint (\ref{eq:c2}). Under Assumption
\ref{asump:LatentIgnorability} and (\ref{eq:2}), it follows 
\begin{multline}
E\left(\frac{1}{n}\sum_{i=1}^{m}\sum_{j=1}^{n_{i}}\frac{A_{ij}}{\hat{e}_{ij}}Y_{ij}\right)=E\left\{ \frac{1}{n}\sum_{i=1}^{m}\sum_{j=1}^{n_{i}}\frac{A_{ij}}{\hat{e}_{ij}}Y_{ij}(1)\right\} \\
=E\left\{ \frac{1}{n}\sum_{i=1}^{m}\sum_{j=1}^{n_{i}}Y_{ij}(1)\right\} =E\{Y(1)\}.\label{eq:3}
\end{multline}
Similarly, we establish 
\begin{equation}
E\left(\frac{1}{n}\sum_{i=1}^{m}\sum_{j=1}^{n_{i}}\frac{1-A_{ij}}{1-\hat{e}_{ij}}Y_{ij}\right)=E\{Y(0)\}.\label{eq:4}
\end{equation}
Combining (\ref{eq:3}) and (\ref{eq:4}), we have $E(\hat{\tau}_{\iptw})\cong E\{Y(1)-Y(0)\}=\tau,$
which yields the asymptotic unbiasedness of $\hat{\tau}_{\iptw}$.
Under certain regularity conditions as in the Appendix, we then have
$\plim_{n\rightarrow\infty}\hat{\tau}_{\iptw}=\tau$. 

It is important to note that to the logistic model is only a working
model. The proposed estimator $\hat{\tau}_{\iptw}$ does not require
specification of this working model to be true. (\ref{eq:c1}) and
(\ref{eq:c2}) play the key role for the unbiasedness of $\hat{\tau}_{\iptw}$.
Therefore, $\hat{\tau}_{\mathrm{\iptw}}$ is robust to specification
of this working propensity score model. 

\begin{theorem}\label{Thm:Asymp Normality}

Suppose that Assumptions \ref{asump:LatentIgnorability} and \ref{asump:positivity},
and the regularity conditions specified in the Appendix hold, and
that the outcome and propensity score follow the generalized linear
mixed effects models in (\ref{eq:glmm}) and (\ref{eq:glmm_ps}).
Suppose further that the number of clusters $m$ and the cluster sample
sizes $n_{i}$, for $i=1,\ldots m$, satisfy the condition that $m\rightarrow\infty$,
as $n\rightarrow\infty$, and $\sup_{1\leq i\leq m}n_{i}=O(n^{1/2})$.
Then, the proposed propensity score weighting estimator in (\ref{eq:proposed estimator}),
subject to constraints (\ref{eq:c1}) and (\ref{eq:c2}), satisfies
\[
V_{1}^{-1/2}(\hat{\tau}_{\iptw}-\tau)\rightarrow\mathcal{N}(0,1),
\]
in distribution, as $n\rightarrow\infty$, where $V_{1}=\mathrm{var}\left(n^{-1}\sum_{i=1}^{m}\sum_{j=1}^{n_{i}}\tau_{ij}\right),$
with $\tau_{ij}=\{\alpha_{ij}(\lambda_{1}^{*},\lambda_{2}^{*})A_{ij}(Y_{ij}-B_{1}^{\T}X_{ij})+B_{1}^{\T}X_{ij}\}-\{\alpha_{ij}(\lambda_{1}^{*},\lambda_{2}^{*})(1-A_{ij})(Y_{ij}-B_{2}^{\T}X_{ij})+B_{2}^{\T}X_{ij}\}$,
\begin{eqnarray*}
B_{1} & = & E\left[\sum_{i=1}^{m}\sum_{j=1}^{n_{i}}\alpha_{ij}(\lambda_{1}^{*},\lambda_{2}^{*})\left\{ 1-\frac{\alpha_{ij}(\lambda_{1}^{*},\lambda_{2}^{*})}{n_{i}}\right\} A_{ij}Y_{ij}X_{ij}^{\T}\right]\\
 &  & \times E\left[\sum_{i=1}^{m}\sum_{j=1}^{n_{i}}\alpha_{ij}(\lambda_{1}^{*},\lambda_{2}^{*})\left\{ 1-\frac{\alpha_{ij}(\lambda_{1}^{*},\lambda_{2}^{*})}{n_{i}}\right\} A_{ij}X_{ij}X_{ij}^{\T}\right]^{-1},\\
B_{2} & = & E\left[\sum_{i=1}^{m}\sum_{j=1}^{n_{i}}\alpha_{ij}(\lambda_{1}^{*},\lambda_{2}^{*})\left\{ 1-\frac{\alpha_{ij}(\lambda_{1}^{*},\lambda_{2}^{*})}{n_{i}}\right\} (1-A_{ij})Y_{ij}X_{ij}^{\T}\right]\\
 &  & \times E\left[\sum_{i=1}^{m}\sum_{j=1}^{n_{i}}\alpha_{ij}(\lambda_{1}^{*},\lambda_{2}^{*})\left\{ 1-\frac{\alpha_{ij}(\lambda_{1}^{*},\lambda_{2}^{*})}{n_{i}}\right\} (1-A_{ij})X_{ij}X_{ij}^{\T}\right]^{-1},
\end{eqnarray*}
and $(\lambda_{1}^{*},\lambda_{2}^{*})^{\T}$ satisfies $E\{\hat{Q}(\lambda_{1}^{*},\lambda_{2}^{*})\}=0$
with $\hat{Q}(\lambda_{1},\lambda_{2})$ defined in (\ref{eq:hat Q}).

\end{theorem} 

The proof is relegated to the Appendix. We now discuss variance estimation.
Let $\hat{\tau}_{ij}=\alpha_{ij}(\hat{\lambda}_{1},\hat{\lambda}_{2})\{A_{ij}(Y_{ij}-\hat{B}_{1}^{\T}X_{ij})-(1-A_{ij})(Y_{ij}-\hat{B}_{2}^{\T}X_{ij})\}+(\hat{B}_{1}-\hat{B}_{2})^{\T}X_{ij}$,
where 
\begin{eqnarray*}
\hat{B}_{1} & = & \sum_{i=1}^{m}\sum_{j=1}^{n_{i}}\alpha_{ij}(\hat{\lambda}_{1},\hat{\lambda}_{2})\left\{ 1-\frac{\alpha_{ij}(\hat{\lambda}_{1},\hat{\lambda}_{2})}{n_{i}}\right\} A_{ij}Y_{ij}X_{ij}^{\T}\\
 &  & \times\left[\sum_{i=1}^{m}\sum_{j=1}^{n_{i}}\alpha_{ij}(\hat{\lambda}_{1},\hat{\lambda}_{2})\left\{ 1-\frac{\alpha_{ij}(\hat{\lambda}_{1},\hat{\lambda}_{2})}{n_{i}}\right\} A_{ij}X_{ij}X_{ij}^{\T}\right]^{-1},\\
\hat{B}_{2} & = & \sum_{i=1}^{m}\sum_{j=1}^{n_{i}}\alpha_{ij}(\hat{\lambda}_{1},\hat{\lambda}_{2})\left\{ 1-\frac{\alpha_{ij}(\hat{\lambda}_{1},\hat{\lambda}_{2})}{n_{i}}\right\} (1-A_{ij})Y_{ij}X_{ij}^{\T}\\
 &  & \times\left[\sum_{i=1}^{m}\sum_{j=1}^{n_{i}}\alpha_{ij}(\hat{\lambda}_{1},\hat{\lambda}_{2})\left\{ 1-\frac{\alpha_{ij}(\hat{\lambda}_{1},\hat{\lambda}_{2})}{n_{i}}\right\} (1-A_{ij})X_{ij}X_{ij}^{\T}\right]^{-1}.
\end{eqnarray*}
Let $\hat{\tau}_{i}=n_{i}^{-1}\sum_{j=1}^{n_{i}}\hat{\tau}_{ij}$
and $\hat{V}_{i}=(n_{i}-1)^{-1}\sum_{j=1}^{n_{i}}(\hat{\tau}_{ij}-\hat{\tau}_{i})^{2}.$
The variance estimator can be constructed as
\[
\hat{V}(\hat{\tau}_{\iptw})=\frac{1}{n}\left(\frac{1}{m-1}\sum_{i=1}^{m}(\hat{\tau}_{i}-\hat{\tau}_{\iptw})^{2}+\frac{1}{m}\sum_{i=1}^{m}\hat{V}_{i}\right).
\]

\section{Extension to clustered survey data\label{sec:Extension-to-survey}}

Clustered data often arise in survey sampling. In complex surveys,
the challenge is to take design information or design weights into
account when developing propensity score methods for causal inference.
In this section, we extend the proposed propensity score weighting
estimator to clustered survey data. Consider a finite population $\mathcal{F}_{N}$
with $M$ clusters and $N_{i}$ units in the $i$th cluster, where
$N=\sum_{i=1}^{M}N_{i}$ denotes the population size. We assume that
in the finite population, $\{A_{ij},X_{ij},Y_{ij}(0),Y_{ij}(1)\}$
follows the superpopulation model $\xi$ as described in Section \ref{sec:Basic-Setup}.
We are interested in estimating the average treatment effect $\tau=E\{Y(1)-Y(0)\}$.

The sample is selected according to a two-stage cluster sampling design.
Specifically, at the first stage, cluster $i$ is sampled with the
first inclusion probability $\pi_{i}$, $i\in S_{I}$, where $S_{I}$
is the index set for the sampled clusters. Let $\pi_{ij}=\mathrm{pr}(i,j\in S_{I})$
be the second inclusion probability for clusters $i$ and $j$ being
sampled. At the second stage, given that cluster $i$ was selected
at the first stage, unit $j$ is sampled with conditional probability
$\pi_{j|i}$, $j=1,\ldots,n_{i}$. The final sample size is $n=\sum_{i\in S_{I}}n_{i}$.
The design weight for unit $j$ in cluster $i$ be $\omega_{ij}=(\pi_{i}\pi_{j|i})^{-1}$,
which reflects the number of units for cluster $i$ in the finite
population this unit $j$ represents. We assume that the design weights
are positive and known throughout the sample. Also, let $\pi_{kl|i}$
be the second inclusion probability for units $k$ and $l$ being
sampled given that cluster $i$ was selected. The second inclusion
probabilities, the $\pi_{ij}$ and $\pi_{kl|i}$'s, are often used
for variance estimation. 

\textcolor{black}{For clustered survey data, }if the propensity score
$e(X_{ij},U_{i})$ is known, we can express the IPTW estimator of
$\tau$\textcolor{black}{{} as} 
\begin{equation}
\hat{\tau}_{\iptw}=\frac{1}{N}\sum_{i\in S_{I}}\sum_{j=1}^{n_{i}}\omega_{ij}\left\{ \frac{A_{ij}Y_{ij}}{e(X_{ij},U_{i})}-\frac{(1-A_{ij})Y_{ij}}{1-e(X_{ij},U_{i})}\right\} .\label{eq:ps estimator}
\end{equation}
Let $E_{\xi}$ and $E_{p}$ denote expectation under the super-population
model and the sampling design, respectively. It is easy to verify
that 
\begin{eqnarray*}
E(\hat{\tau}_{\iptw}) & = & E_{\xi}\{E_{p}(\hat{\tau}_{\iptw})\}\\
 & = & E_{\xi}\left[\frac{1}{N}\sum_{i=1}^{M}\sum_{j=1}^{N_{i}}\left\{ \frac{A_{ij}Y_{ij}}{e(X_{ij},U_{i})}-\frac{(1-A_{ij})Y_{ij}}{1-e(X_{ij},U_{i})}\right\} \right]=\tau.
\end{eqnarray*}

In practice, since the propensity score $e(X_{ij},U_{i})$ is often
unknown, (\ref{eq:ps estimator}) is not infeasible. To estimate the
propensity score, we now require the propensity score estimate $\hat{e}_{ij}$
satisfy the following design-weighted moment constraints 
\begin{eqnarray}
\sum_{i\in S_{I}}\sum_{j=1}^{n_{i}}\omega_{ij}\frac{A_{ij}}{\hat{e}_{ij}}X_{ij} & = & \sum_{i\in S_{I}}\sum_{j=1}^{n_{i}}\omega_{ij}\frac{1-A_{ij}}{1-\hat{e}_{ij}}X_{ij}=\sum_{i\in S_{I}}\sum_{j=1}^{n_{i}}\omega_{ij}X_{ij},\label{eq:c1*}\\
\sum_{j=1}^{n_{i}}\omega_{ij}\frac{A_{ij}}{\hat{e}_{ij}} & = & \sum_{j=1}^{n_{i}}\omega_{ij}\frac{1-A_{ij}}{1-\hat{e}_{ij}}=N_{i},\quad(i\in S_{I}).\label{eq:c3*}
\end{eqnarray}
These moment constraints (\ref{eq:c1*}) and (\ref{eq:c3*}) are the
sample version of (\ref{eq:p1}) and (\ref{eq:p2}), respectively.

To obtain the propensity score estimate that achieves (\ref{eq:c1*})
and (\ref{eq:c3*}), we use the calibration technique in the following
steps:
\begin{description}
\item [{Step$\ $1.}] \label{step1-1} Obtain an initial propensity score
estimate $\hat{e}_{ij}^{0}$ under some working propensity score model,
e.g. a logistic linear model fitted to $(A_{ij},X_{ij})$, each unit
weighted by the design weight $\omega_{ij}$. This in turn provides
an initial set of inverse propensity score weights, $\mathbb{W}^{0}=\{d_{ij};i=1,\ldots,m,j=1,\ldots,n_{i}\}$,
where $d_{ij}=1/e_{ij}^{0}$ if $A_{ij}=1$ and $d_{ij}=1/(1-e_{ij}^{0})$
if $A_{ij}=0$.
\item [{Step$\ $2.}] \label{step2-1} Modify the initial set of weights
$\mathbb{W}^{0}$ to a new set of weights $\mathbb{W}=\{\alpha_{ij};i=1,\ldots,m,j=1,\ldots,n_{i}\}$
by minimizing $\sum_{i=1}^{m}\sum_{j=1}^{n_{i}}\omega_{ij}\alpha_{ij}\log(\alpha_{ij}/d_{ij}),$
subject to (\ref{eq:c1*}) and (\ref{eq:c3*}). By Lagrange Multiplier,
$\alpha_{ij}$ can be obtained as
\begin{multline*}
\alpha_{ij}(\lambda_{1},\lambda_{2})=\frac{N_{i}A_{ij}d_{ij}\exp(\lambda_{1}X_{ij}A_{ij})}{\sum_{j=1}^{n_{i}}\omega_{ij}A_{ij}d_{ij}\exp(\lambda_{1}X_{ij}A_{ij})}\\
+\frac{N_{i}(1-A_{ij})d_{ij}\exp\{\lambda_{2}X_{ij}(1-A_{ij})\}}{\sum_{j=1}^{n_{i}}\omega_{ij}(1-A_{ij})d_{ij}\exp\{\lambda_{2}X_{ij}(1-A_{ij})\}},
\end{multline*}
where $(\lambda_{1},\lambda_{2})^{\T}$ is the solution to the following
equation
\begin{multline}
\hat{Q}(\lambda_{1},\lambda_{2})=\left(\begin{array}{c}
\hat{Q}_{1}(\lambda_{1},\lambda_{2})\\
\hat{Q}_{2}(\lambda_{1},\lambda_{2})
\end{array}\right)\\
=\left(\begin{array}{c}
N^{-1}\sum_{i\in S_{I}}\sum_{j=1}^{n_{i}}\omega_{ij}\left\{ A_{ij}\alpha_{ij}(\lambda_{1},\lambda_{2})-1\right\} X_{ij}\\
N^{-1}\sum_{i\in S_{I}}\sum_{j=1}^{n_{i}}\omega_{ij}\left\{ (1-A_{ij})\alpha_{ij}(\lambda_{1},\lambda_{2})-1\right\} X_{ij}
\end{array}\right)=0.\label{eq:hat Q-1}
\end{multline}
\item [{Step$\ $3.}] Obtain the propensity score estimate as $\hat{e}_{ij}=\alpha_{ij}(\hat{\lambda}_{1},\hat{\lambda}_{2})^{-A_{ij}}\{1-\alpha_{ij}(\hat{\lambda}_{1},\hat{\lambda}_{2})\}^{-1+A_{ij}}$.
\end{description}
Finally, our proposed IPTW estimator is 
\begin{equation}
\hat{\tau}_{\mathrm{\iptw}}=\frac{1}{N}\sum_{i\in S_{I}}\sum_{j=1}^{n_{i}}\omega_{ij}\left\{ \frac{A_{ij}Y_{ij}}{\hat{e}_{ij}}-\frac{(1-A_{ij})Y_{ij}}{1-\hat{e}_{ij}}\right\} .\label{eq:survey ps estimator}
\end{equation}
In the above procedure, the design wights are used in both the propensity
score estimates and the weighting estimator. 

We now consider the asymptotic property of $\hat{\tau}_{\iptw}$ in
(\ref{eq:survey ps estimator}). We use an asymptotic framework, where
the sample size $n$ indexes a sequence of finite populations and
samples (\citealp{fuller2009sampling}; Section $1.3$), such that
the population size $N$ increases with $n$, but the cluster sample
sizes $N_{i}$ may remain small. In addition, we have the following
regularity conditions for the sampling mechanism.

\begin{assumption}\label{asmp:sampling} (i) The first-order inclusion
probability $\pi_{i}\pi_{j|i}$ is positive and uniformly bounded
in the sense that there exist positive constants $C_{1}$ and $C_{2}$
that do not depend on $N$, such that $C_{1}\le\pi_{i}\pi_{j|i}Nn^{-1}\le C_{2},$
for any $i$ and $j$; (ii) the sequence of Hotvitz-Thompson estimators
$\hat{Y}_{\HT}=N^{-1}\sum_{i\in S_{I}}\sum_{j=1}^{n_{i}}\omega_{ij}y_{i}$
satisfies $\var_{p}(\hat{Y}_{\HT})=O\left(n^{-1}\right)$ and $\left\{ \var_{p}(\hat{Y}_{\HT})\right\} ^{-1/2}(\hat{Y}_{\HT}-\bar{Y})\mid\mathcal{F}_{N}\rightarrow\mathcal{N}(0,1),$
in distribution, as $n\rightarrow\infty$, where $\bar{Y}=N^{-1}\sum_{i=1}^{M}\sum_{j=1}^{N_{i}}y_{i}$
is the population mean of $Y$, and the reference distribution is
the randomization distribution generated by the sampling mechanism.
\end{assumption} Sufficient conditions for the asymptotic normality
of the Hotvitz-Thompson estimators are discussed in Chapter 1 of \citet{fuller2009sampling}.

\begin{theorem}\label{Thm:Asymp Normality-1}

Suppose that Assumptions \ref{asump:LatentIgnorability} and \ref{asump:positivity},
and the regularity conditions specified in the Appendix hold, and
that the outcome and propensity score follow generalized linear mixed
effects models in (\ref{eq:glmm}) and (\ref{eq:glmm_ps}). Suppose
further that the sequence of finite populations and samples satisfy
Assumption \ref{asmp:sampling}. Then, the proposed propensity score
weighting estimator in (\ref{eq:survey ps estimator}), subject to
constraints (\ref{eq:c1*}) and (\ref{eq:c3*}), satisfies 
\[
V_{2}^{-1}(\hat{\tau}_{\iptw}-\tau)\rightarrow\mathcal{N}(0,1),
\]
in distribution, as $n\rightarrow\infty$, where $V_{2}=\mathrm{var}\left(N^{-1}\sum_{i\in S_{I}}\sum_{j=1}^{n_{i}}\omega_{ij}\tau_{ij}\right),$
with $\tau_{ij}=\{\alpha_{ij}(\lambda_{1}^{*},\lambda_{2}^{*})A_{ij}(Y_{ij}-B_{1}^{\T}X_{ij})+B_{1}^{\T}X_{ij}\}-\{\alpha_{ij}(\lambda_{1}^{*},\lambda_{2}^{*})(1-A_{ij})(Y_{ij}-B_{2}^{\T}X_{ij})+B_{2}^{\T}X_{ij}\}$,
\begin{eqnarray*}
B_{1} & = & E\left[\sum_{i=1}^{M}\sum_{j=1}^{N_{i}}\alpha_{ij}(\lambda_{1}^{*},\lambda_{2}^{*})\left\{ 1-\frac{\alpha_{ij}(\lambda_{1}^{*},\lambda_{2}^{*})}{n_{i}}\right\} A_{ij}Y_{ij}X_{ij}^{\T}\right]\\
 &  & \times E\left[\sum_{i=1}^{M}\sum_{j=1}^{N_{i}}\alpha_{ij}(\lambda_{1}^{*},\lambda_{2}^{*})\left\{ 1-\frac{\alpha_{ij}(\lambda_{1}^{*},\lambda_{2}^{*})}{n_{i}}\right\} A_{ij}X_{ij}X_{ij}^{\T}\right]^{-1},\\
B_{2} & = & E\left[\sum_{i=1}^{M}\sum_{j=1}^{N_{i}}\alpha_{ij}(\lambda_{1}^{*},\lambda_{2}^{*})\left\{ 1-\frac{\alpha_{ij}(\lambda_{1}^{*},\lambda_{2}^{*})}{n_{i}}\right\} (1-A_{ij})Y_{ij}X_{ij}^{\T}\right]\\
 &  & \times E\left[\sum_{i=1}^{M}\sum_{j=1}^{N_{i}}\alpha_{ij}(\lambda_{1}^{*},\lambda_{2}^{*})\left\{ 1-\frac{\alpha_{ij}(\lambda_{1}^{*},\lambda_{2}^{*})}{n_{i}}\right\} (1-A_{ij})X_{ij}X_{ij}^{\T}\right]^{-1},
\end{eqnarray*}
and $(\lambda_{1}^{*},\lambda_{2}^{*})^{\T}$ satisfies $E\{\hat{Q}(\lambda_{1}^{*},\lambda_{2}^{*})\}=0$
with $\hat{Q}(\lambda_{1},\lambda_{2})$ defined in (\ref{eq:hat Q-1}). 

\end{theorem} 

For variance estimation of $\hat{\tau}_{\iptw}$, let $\hat{\tau}_{ij}=\alpha_{ij}(\hat{\lambda}_{1},\hat{\lambda}_{2})\{A_{ij}(Y_{ij}-\hat{B}_{1}^{\T}X_{ij})-(1-A_{ij})(Y_{ij}-\hat{B}_{2}^{\T}X_{ij})\}+(\hat{B}_{1}-\hat{B}_{2})^{\T}X_{ij}$,
where 
\begin{eqnarray*}
\hat{B}_{1} & = & \sum_{i\in S_{I}}\sum_{j=1}^{n_{i}}\omega_{ij}\alpha_{ij}(\hat{\lambda}_{1},\hat{\lambda}_{2})\left\{ 1-\frac{\alpha_{ij}(\hat{\lambda}_{1},\hat{\lambda}_{2})}{n_{i}}\right\} A_{ij}Y_{ij}X_{ij}^{\T}\\
 &  & \times\left[\sum_{i\in S_{I}}\sum_{j=1}^{n_{i}}\omega_{ij}\alpha_{ij}(\hat{\lambda}_{1},\hat{\lambda}_{2})\left\{ 1-\frac{\alpha_{ij}(\hat{\lambda}_{1},\hat{\lambda}_{2})}{n_{i}}\right\} A_{ij}X_{ij}X_{ij}^{\T}\right]^{-1},\\
\hat{B}_{2} & = & \sum_{i\in S_{I}}\sum_{j=1}^{n_{i}}\omega_{ij}\alpha_{ij}(\hat{\lambda}_{1},\hat{\lambda}_{2})\left\{ 1-\frac{\alpha_{ij}(\hat{\lambda}_{1},\hat{\lambda}_{2})}{n_{i}}\right\} (1-A_{ij})Y_{ij}X_{ij}^{\T}\\
 &  & \times\left[\sum_{i\in S_{I}}\sum_{j=1}^{n_{i}}\omega_{ij}\alpha_{ij}(\hat{\lambda}_{1},\hat{\lambda}_{2})\left\{ 1-\frac{\alpha_{ij}(\hat{\lambda}_{1},\hat{\lambda}_{2})}{n_{i}}\right\} (1-A_{ij})X_{ij}X_{ij}^{\T}\right]^{-1}.
\end{eqnarray*}
Let $\hat{\tau}_{i}=\sum_{j=1}^{n_{i}}\pi_{j|i}^{-1}\hat{\tau}_{ij}$
and 
\[
\hat{V}_{i}=\sum_{k=1}^{n_{i}}\sum_{l=1}^{n_{i}}\frac{\pi_{kl|i}-\pi_{k|i}\pi_{l|i}}{\pi_{kl|i}}\frac{\hat{\tau}_{ik}}{\pi_{k|i}}\frac{\hat{\tau}_{il}}{\pi_{l|i}}.
\]
The variance estimator can be constructed as
\[
\hat{V}(\hat{\tau}_{\mathrm{\iptw}})=\frac{1}{N^{2}}\left(\sum_{i\in S_{I}}\sum_{j\in S_{I}}\frac{\pi_{ij}-\pi_{i}\pi_{j}}{\pi_{ij}}\frac{\hat{\tau}_{i}}{\pi_{i}}\frac{\hat{\tau}_{j}}{\pi_{j}}+\sum_{i\in S_{I}}\frac{\hat{V}_{i}}{\pi_{i}}\right).
\]

\section{A simulation study }

We conduct simulation studies to evaluate the finite-sample performance
of the proposed estimator. We first generate finite populations and
then select a sample from each finite population using a two-stage
cluster sampling design. 

In the first setting, we specify the number of clusters in the population
to be $M=10,000$, and the size of the $i$th cluster size $N_{i}$
to be the integer part of $500\times\exp(2+U_{i})/\{1+\exp(2+U_{i})\}$,
where $U_{i}\sim\N(0,1)$. The cluster sizes range from $100$ to
$500$. The potential outcomes are generated according to linear mixed
effects models, $Y_{ij}(0)=X_{ij}+U_{i}+e_{ij}$ and $Y_{ij}(1)=X_{ij}+\tau+\tau U_{i}+e_{ij}$,
where $\tau=2$, $X_{ij}\sim\N(0,1)$, $e_{ij}\sim\N(0,1)$, $U_{i}$,
$X_{ij}$, and $e_{ij}$ are independent, for $i=1,\ldots,M$, $j=1,\ldots,N_{i}$.
The parameter of interest is $\tau$. We consider three propensity
score models, $\mathrm{pr}(A_{ij}=1\mid X_{ij};U_{i})=h(\gamma_{0}+\gamma_{1}U_{i}+X_{ij})$,
with $h(\cdot)$ being the inverse logit, probit and complementary
log-log link function, for generating $A_{ij}$. We set $(\gamma_{0},\gamma_{1})$
to be $(-0.5,1)$, $(-0.25,0.5)$ and $(-0.5,0.1)$ for the above
three propensity score models, respectively. The observed outcome
is $Y_{ij}=A_{ij}Y_{ij}(1)+(1-A_{ij})Y_{ij}(0)$. From each realized
population, $m$ clusters are sampled by Probability-Proportional-to-Size
(PPS) sampling with the measure of size $N_{i}$. So the first-order
inclusion probability of selecting cluster $i$ is equal to $\pi_{i}=mN_{i}/\sum_{i=1}^{m}N_{i}$,
which implicitly depends on the unobserved random effect. Once the
clusters are sampled, the $n_{i}$ units in the $i$th selected cluster
are sampled by Poison sampling with the corresponding first-order
inclusion probability $\pi_{j|i}=n_{e}z_{ij}/(\sum_{j=1}^{M_{i}}z_{ij})$,
where $z_{ij}=0.5$ if $e_{ij}<0$ and 1 if $e_{ij}>0$. With this
sampling design, the units with $e_{ij}>0$ are sampled with a chance
twice as big as the units with $e_{ij}<0$. We consider three combinations
of $m$ and $n_{e}$: (i) $(m,n_{e})=(50,50)$; (ii) $(m,n_{e})=(100,30)$,
representing a large number of small clusters; and (iii) $(m,n_{e})=(30,100)$,
representing a small number of large clusters. 

In the second setting, all data-generating mechanisms are the same
as in the first setting, except that the potential outcomes are generated
according to logistic linear mixed effects models, $Y_{ij}(0)\sim\mathrm{Bernoulli}(p_{ij}^{0})$
with $\mathrm{logit}(p_{ij}^{0})=X_{ij}+U_{i}$ and $Y_{ij}(1)\sim\mathrm{Bernoulli}(p_{ij}^{1})$
with $\mathrm{logit}(p_{ij}^{1})=X_{ij}+\tau+\tau u_{i}$. Moreover,
in the 2-stage cluster sampling, $\pi_{j|i}=n_{e}z_{ij}/(\sum_{j=1}^{M_{i}}z_{ij})$,
where $z_{ij}=0.5$ if $Y_{ij}=0$ and 1 if $Y_{ij}=1$. With this
sampling design, the units with $Y_{ij}=1$ are sampled with a chance
twice as big as the units with $Y_{ij}=0$. 

We compare four estimators for $\tau$: (i) $\hat{\tau}_{\mathrm{simp}}$,
the simple design-weighted estimator without propensity score adjustment;
(ii) $\hat{\tau}_{\mathrm{fix}}$, the weighting estimator in (\ref{eq:iptw})
with the propensity score estimated by a logistic linear fixed effects
model with a cluster-level main effect; (iii) $\hat{\tau}_{\mathrm{ran}}$,
the weighting estimator in (\ref{eq:iptw}) with the propensity score
estimated by a logistic linear mixed effects model where the cluster
effect is random; (iv) $\hat{\tau}_{\mathrm{\iptw}}$, the proposed
estimator with calibrations (\ref{eq:c1*}) and (\ref{eq:c3*}). Table
\ref{tab:1} shows biases, variances and coverages for $95\%$ confidence
intervals from $1,000$ simulated data sets. The simple estimator
shows large biases across difference scenarios, even adjusting for
sampling design. This suggests that the covariate distributions are
different between the treatment groups in the finite population, contributing
to the bias. $\hat{\tau}_{\mathrm{fix}}$ works well under Scenario
1 with the linear mixed effects model for the outcome and the logistic
linear mixed effects model for the propensity score; however, its
performance is not satisfactory in other scenarios. Moreover, $\hat{\tau}_{\mathrm{fix}}$
shows the largest variance among the four estimators in most of scenarios.
This is because for a moderate or large number of clusters, there
are too many free parameters, and hence the propensity score estimates
may not be stable. For $\hat{\tau}_{\mathrm{ran}}$, we assume that
the cluster effect is random, which reduces the number of free parameters.
As a result, $\hat{\tau}_{\mathrm{ran}}$ shows less variability than
$\hat{\tau}_{\mathrm{fix}}$. Nonetheless, both $\hat{\tau}_{\mathrm{fix}}$
and $\hat{\tau}_{\mathrm{ran}}$ can not control the bias well. The
proposed estimator shows small biases and good empirical coverages
across all scenarios. Notably, to compute $\hat{\tau}_{\iptw},$ we
used a working model, a logistic linear model, to provide an initial
set of weights. When the true propensity score is probit or complementary
log-log model, $\hat{\tau}_{\mathrm{\iptw}}$ still has small biases.
This suggests that our proposed estimator is robust to the working
model in our simulation settings. 

\begin{table}
\caption{\label{tab:1}Simulation results: bias, variance ($\times10^{-3}$)
and coverage ($\%$) of $95\%$ confidence intervals based on $1,000$
Monte Carlo samples; the outcome is linear and logistic linear mixed
effects model and the propensity score is logistic, probit or complementary
log-log (C-loglog).}

\centering{}%
\begin{tabular}{cccccccccccc}
\hline 
 & \multicolumn{3}{c}{$(m,n_{e})=(50,50)$} &  & \multicolumn{3}{c}{$(m,n_{e})=(100,30)$} &  & \multicolumn{3}{c}{$(m,n_{e})=(30,100)$}\tabularnewline
Method & bias & var & cvg &  & bias & var & cvg &  & bias & var & cvg\tabularnewline
\hline 
\multicolumn{12}{c}{Scenario 1: Linear outcome \& Logistic propensity score}\tabularnewline
$\hat{\tau}_{\mathrm{simp}}$ & \textcolor{black}{-0.37} & 22 & 27.4  &  & -0.38 & 12 & 8.7 &  & -0.38 & 35 & 42.3\tabularnewline
$\hat{\tau}_{\mathrm{fix}}$ & -0.01 & \textcolor{black}{36} & \textcolor{black}{95.6} &  & \textcolor{black}{0.00} & \textcolor{black}{21} & \textcolor{black}{95.6} &  & \textcolor{black}{-0.01} & \textcolor{black}{42} & \textcolor{black}{95.2}\tabularnewline
$\hat{\tau}_{\mathrm{ran}}$ & 0.14 & \textcolor{black}{26} & \textcolor{black}{90.2} &  & \textcolor{black}{0.21} & \textcolor{black}{14} & \textcolor{black}{64.6} &  & \textcolor{black}{0.07} & \textcolor{black}{37} & 94.7\tabularnewline
$\hat{\tau}_{\mathrm{cal}}$ & 0.01 & 26 & 94.5 &  & 0.02 & 11 & 95.1 &  & 0.00 & 33 & 95.6\tabularnewline
\hline 
\multicolumn{12}{c}{Scenario 2: Linear outcome \& Probit propensity score}\tabularnewline
$\hat{\tau}_{\mathrm{simp}}$ & -0.29 & 16 & 34.4 &  & -0.08 & 9 & \textcolor{black}{2.3} &  & -0.22 & 30 & 65.6\tabularnewline
$\hat{\tau}_{\mathrm{fix}}$ & 0.08 & 35 & 90.3 &  & -0.10 & 19 & \textcolor{black}{4.5} &  & 0.12 & 69 & 90.4\tabularnewline
$\hat{\tau}_{\mathrm{ran}}$ & 0.24 & 28 & 73.9 &  & -0.07 & 16 & \textcolor{black}{29.9} &  & 0.21 & 60 & 85.5\tabularnewline
$\hat{\tau}_{\mathrm{cal}}$ & 0.01 & 22 & 94.9 &  & 0.01 & 11 & \textcolor{black}{95.4} &  & 0.00 & 33 & 94.6\tabularnewline
\hline 
\multicolumn{12}{c}{Scenario 3: Linear outcome \& C-loglog propensity score}\tabularnewline
$\hat{\tau}_{\mathrm{simp}}$ & -0.21 & 20 & 62.0 &  & -0.21 & 10 & 41.2 &  & -0.22 & 30 & 65.6\tabularnewline
$\hat{\tau}_{\mathrm{fix}}$ & 0.12 & 48 & 88.8 &  & 0.12 & 36 & 82.7 &  & 0.12 & 69 & 90.4\tabularnewline
$\hat{\tau}_{\mathrm{ran}}$ & 0.29 & 38 & 69.1 &  & 0.36 & 22 & 32.5 &  & 0.21 & 60 & 85.5\tabularnewline
$\hat{\tau}_{\mathrm{cal}}$ & 0.00 & 21 & 95.3 &  & 0.00 & 10 & 95.1 &  & 0.00 & 33 & 94.6\tabularnewline
\hline 
\multicolumn{12}{c}{Scenario 4: Logistic outcome \& Logistic propensity score}\tabularnewline
$\hat{\tau}_{\mathrm{simp}}$ & -0.11 & 100 & 9.1 &  & -0.11 & 540 & 0.5 &  & -0.11 & 160 & 20.5 \tabularnewline
$\hat{\tau}_{\mathrm{fix}}$ & -0.11 & \textcolor{black}{44} & 0.3 &  & -0.11 & 38 & 0.1 &  & -0.11 & 39 & 0.1\tabularnewline
$\hat{\tau}_{\mathrm{ran}}$ & -0.09 & \textcolor{black}{33} & 1.3  &  & -0.08 & 21 & 0.5 &  & -0.10 & 34 & 0.3\tabularnewline
$\hat{\tau}_{\mathrm{cal}}$ & 0.01 & 74  & 96.3  &  & 0.01 & 55 & 95.2 &  & 0.01 & 74 & 95.9 \tabularnewline
\hline 
\multicolumn{12}{c}{Scenario 5: Logistic outcome \& Probit propensity score}\tabularnewline
$\hat{\tau}_{\mathrm{simp}}$ & -0.08 & 58 & 13.1  &  & -0.08 & 34 & 2.3 &  & -0.08 & 81 & 25.3\tabularnewline
$\hat{\tau}_{\mathrm{fix}}$ & -0.10  & 93 & 6.9 &  & -0.10  & 85 & 4.5 &  & -0.10  & 73 & 3.8\tabularnewline
$\hat{\tau}_{\mathrm{ran}}$ & -0.08 & 67 & 23.0  &  & -0.07 & 48 & 29.9 &  & -0.09 & 61 & 8.3\tabularnewline
$\hat{\tau}_{\mathrm{cal}}$ & 0.01 & 89  & 94.7 &  & 0.01 & 65 & 95.4 &  & 0.01 & 84 & 95.0\tabularnewline
\hline 
\multicolumn{12}{c}{Scenario 6: Logistic outcome \& C-loglog propensity score}\tabularnewline
$\hat{\tau}_{\mathrm{simp}}$ & -0.06 & 0.3 & 3.2 &  & -0.06 & 0.2 & 1.0 &  & -0.06 & 0.2 & 3.7\tabularnewline
$\hat{\tau}_{\mathrm{fix}}$ & -0.05 & 0.5 & 44.6 &  & -0.05 & 0.5 & 43.6 &  & -0.05 & 0.5 & 43.0\tabularnewline
$\hat{\tau}_{\mathrm{ran}}$ & -0.03 & 0.5 & 95.4 &  & -0.03 & 0.4 & 97.3 &  & -0.03 & 0.4 & 92.8\tabularnewline
$\hat{\tau}_{\mathrm{cal}}$ & -0.01 & 0.7 & 95.5 &  & 0.00 & 0.6 & 95.8 &  & -0.01 & 0.7 & 95.2\tabularnewline
\hline 
\end{tabular}
\end{table}

\section{An Application}

We examine the 2007\textendash 2010 BMI surveillance data from Pennsylvania
Department of Health to investigate the effect of School Body Mass
Index Screening (SBMIS) on the annual overweight and obesity prevalence
in elementary schools in Pennsylvania. Early studies have shown that
SBMIS has been associated with increased parental awareness of child
weight \citep{harris2009effect,ebbeling2012randomized}. However,
there have been mixed findings about the effect of screening on reducing
prevalence of overweight and obesity \citep{harris2009effect,thompson2009arkansas}.
The data set includes $493$ school districts in Pennsylvania. The
baseline is the school year 2007. The schools are clustered by two
factors: location (rural, suburban, and urban), and population density
(low, median, and high). This results in five clusters: rural-low,
rural-median, rural-high, suburban-high, and urban-high. Let $A=1$
if the school implemented SBMIS, and $A=0$ if the school did not.
In this data set, $63\%$ of schools implemented SBMIS, and the percentages
of schools implemented SBMIS across the clusters range from $45\%$
to $70\%$, indicating cluster-level heterogeneity of treatment. The
outcome variable $Y$ is the annual overweight and obesity prevalence
for each school district in the school year 2010. The prevalence is
calculated by dividing the number with BMI$>85$th by the total number
of students screened for each school district. For each school, we
obtain school characteristics including the baseline prevalence of
overweight and obesity ($X_{1}$), and percentage of reduced and free
lunch ($X_{2}$). 

For a direct comparison, the average difference of the prevalence
of overweight and obesity for schools that implemented SBMIS and those
that did not is $8.78\%$. This unadjusted difference in the prevalence
of overweight and obesity ignores differences in schools and clusters.
\textcolor{black}{To take the cluster-level heterogeneity of treatment
into account,} we consider three propensity score models: (i) a logistic
linear fixed effects model with linear predictors including $X_{1}$,
$X_{2}$, and a fixed intercept for each cluster; (ii) a logistic
linear mixed effects model with linear predictors including fixed
effects $X_{1}$, $X_{2}$, and a random effect for each cluster;
(iii) the proposed calibrated propensity score. Using the estimated
propensity score, we estimate $\tau=E\{Y(1)-Y(0)\}$ by the weighting
method. 

Table \ref{tab:Results-3} displays the standardized differences of
means for $X_{1}$ and $X_{2}$ between the treated and control groups
for each cluster and the whole population, standardized by the standard
errors in the whole population. Without any adjustment, there are
large differences in means for $X_{1}$ and $X_{2}$. For this specific
data set, the three methods for modeling and estimating the propensity
score are similar in balancing the covariate distributions between
the treated and control groups. All three propensity score weighting
methods improve the balance for $X_{1}$ and $X_{2}$. Table \ref{tab:Results-2}
displays point estimates and variance estimates based on $500$ bootstrap
replicates. The simple estimator shows that the screening has a significant
effect in reducing the prevalence of overweight and obesity. However,
this may be due to confounders. After adjusting for the confounders,
the screening does not have a significant effect. Given the different
sets of assumptions for the different methods, this conclusion is
reassuring.

\begin{table}
\caption{\label{tab:Results-3}Balance Check }

\centering{}%
\begin{tabular}{cccccc}
\hline 
 &  & simple & fixed  & random & calibration\tabularnewline
\hline 
 & Cluster 1  & 1.68 & -0.22 & 0.68 & 0.20\tabularnewline
 & Cluster 2 & 1.21 & 0.10 & -0.41 & 0.10\tabularnewline
$X_{1}$ & Cluster 3  & 1.75 & -0.02 & 0.99 & 0.02\tabularnewline
 & Cluster 4  & 0.86 & -0.04 & -1.05 & 0.02\tabularnewline
 & Cluster 5  & -0.36 & 0.37 & -1.39 & 0.33\tabularnewline
 & Whole Pop  & 1.28 & -0.02 & -0.02 & 0\tabularnewline
\hline 
 & Cluster 1  & 0.48 & 0.02 & 0.30 & 0.03\tabularnewline
 & Cluster 2 & 0.43 & 0.13 & -0.01 & 0.14\tabularnewline
$X_{2}$ & Cluster 3  & 0.73 & 0.01 & 0.46 & 0.02\tabularnewline
 & Cluster 4  & 0.18 & -0.08 & -0.34 & -0.07\tabularnewline
 & Cluster 5  & -0.57 & -0.39 & -1.53 & -0.44\tabularnewline
 & Whole Pop  & 0.39 & -0.003 & -0.001 & 0\tabularnewline
\hline 
\end{tabular}
\end{table}

\begin{table}
\caption{\label{tab:Results-2}Results: estimate, variance estimate (ve) based
on $500$ bootstrap replicates, and $95\%$ confidence interval (c.i.)}

\centering{}%
\begin{tabular}{cccc}
\hline 
 & estimate & ve & $95\%$ c.i.\tabularnewline
\hline 
simple & 8.78 & 2.11 & (5.94, 11.63)\tabularnewline
fixed & 0.47 & 0.44 & (-0.83, 1.77)\tabularnewline
random & 0.52 & 0.44 & (-0.77, 1.82)\tabularnewline
calibration & 0.53 & 0.39 & (-0.71, 1.76)\tabularnewline
\hline 
\end{tabular}
\end{table}

\section{Discussion}

The IPTW estimator is not efficient in general. Semiparametric efficiency
bounds for estimating the average treatment effects in the setting
with i.i.d. random variables were derived by \citet{hahn1998role}.
He showed that the efficient influence function for the average treatment
effect depends on both the propensity score and the outcome model.
An important implication is that combining the propensity score model
and the outcome regression model can improve efficiency of the IPTW
estimator. For clustered data, since the data are correlated through
the random cluster effects, the efficiency theory established for
the i.i.d. data is not applicable. It remains an interesting avenue
for future research to develop the semiparametric efficiency theory
for clustered data.

In this article, we assumed that there is no interference between
units. This setup is not uncommon. In our application, the treatment
was implemented school-wise. The potential outcomes for one school
are likely to be unaffected by the treatments implemented at other
schools, and therefore the assumption of no interference is likely
to hold. However, in other settings, this assumption may not hold.
A classical example is given in infectious diseases \citep{ross1916application,hudgens2008toward},
where whether one person becomes infected depends on who else in the
population is vaccinated. Extension of our calibration estimation
to take the interference structure into account in these settings
is also an interesting topic for future research. 

\section*{Acknowledgments }

We would like to thank Peng Ding and Fan Li for insightful and fruitful
discussions.

\section*{Appendix }

\global\long\def\theequation{A\arabic{equation}}
 \setcounter{equation}{0} 

\global\long\def\thesection{A\arabic{section}}
 \setcounter{equation}{0} 

\global\long\def\thetable{A\arabic{table}}
 \setcounter{equation}{0} 

\global\long\def\theexample{A\arabic{example}}
 \setcounter{equation}{0} 

\global\long\def\thetheorem{A\arabic{theorem}}
 \setcounter{equation}{0} 

\global\long\def\thecondition{A\arabic{condition}}
 \setcounter{equation}{0} 

\global\long\def\theremark{A\arabic{remark}}
 \setcounter{equation}{0} 

\global\long\def\thestep{A\arabic{step}}
 \setcounter{equation}{0} 

\global\long\def\theassumption{A\arabic{assumption}}
 \setcounter{equation}{0} 

\global\long\def\theproof{A\arabic{proof}}
 \setcounter{equation}{0} 

\subsection*{Appendix A. Regularity conditions}

\begin{condition}\label{RegCondition} (i) $\hat{\tau}_{\iptw}(\lambda_{1},\lambda_{2})\rightarrow\tau$
in probability uniformly in a compact set $\mathcal{B}$ as $n\rightarrow\infty$;
(ii) $\hat{Q}(\lambda_{1},\lambda_{2})\rightarrow Q(\lambda_{1},\lambda_{2})=E\{\hat{Q}(\lambda_{1},\lambda_{2})\}$
in probability uniformly in $\mathcal{B}$ as $n\rightarrow\infty$,
and there exists a unique $(\lambda_{1}^{*},\lambda_{2}^{*})\in\mathcal{B}$
such that $Q(\lambda_{1}^{*},\lambda_{2}^{*})=0$; (iii) $\partial\hat{\tau}_{\iptw}(\lambda_{1},\lambda_{2})/\partial(\lambda_{1},\lambda_{2})^{\T}$
and $\partial\hat{Q}(\lambda_{1},\lambda_{2})/\partial(\lambda_{1},\lambda_{2})^{\T}$
are continuous at $(\lambda_{1},\lambda_{2})$ in $\mathcal{B}$ almost
surely; (iv)$E||X_{ij}||^{3}<\infty$, $E|Y_{ij}(0)|^{3}<\infty$,
and $E|Y_{ij}(1)|^{3}<\infty$; (v) the matrix $E\left\{ \partial\hat{Q}(\lambda_{1}^{*},\lambda_{2}^{*})/\partial(\lambda_{1},\lambda_{2})^{\T}\right\} $
is invertible. 

\end{condition}

Conditions \ref{RegCondition} (i)\textendash (iii) ensure that $\plim_{n\rightarrow\infty}(\hat{\lambda}_{1},\hat{\lambda}_{2})=(\lambda_{1}^{*},\lambda_{2}^{*})$
and 
\[
\plim_{n\rightarrow\infty}\hat{\tau}_{\iptw}(\hat{\lambda}_{1},\hat{\lambda}_{2})=\tau,
\]
which is similar to Corollary II.2 of \citet{andersen1982cox}. Condition
\ref{RegCondition} (iv) is a moment condition for the central limit
theorem.

\subsection*{Appendix B. Proof of Theorem 1}

Write $\hat{\tau}_{\iptw}(\lambda_{1},\lambda_{2})=n^{-1}\sum_{i=1}^{m}\sum_{j=1}^{n_{i}}\alpha_{ij}(\lambda_{1},\lambda_{2})Y_{ij},$
where $\alpha_{ij}(\lambda_{1},\lambda_{2})$ is defined in (\ref{eq:alpha_ij}).
The proposed estimator is $\hat{\tau}_{\iptw}(\hat{\lambda}_{1},\hat{\lambda}_{2})$,
where $(\hat{\lambda}_{1},\hat{\lambda}_{2})$ satisfies $\hat{Q}(\hat{\lambda}_{1},\hat{\lambda}_{2})=0$,
where $\hat{Q}(\lambda_{1},\lambda_{2})$ is defined in (\ref{eq:hat Q}).
Let $(\lambda_{1}^{*},\lambda_{2}^{*})$ satisfy $E\{Q(\lambda_{1}^{*},\lambda_{2}^{*})\}=0$. 

Under Conditions \ref{RegCondition}, using the standard linearization
technique, we obtain 
\begin{eqnarray*}
\hat{\tau}_{\mathrm{\iptw}}(\hat{\lambda}_{1},\hat{\lambda}_{2}) & \cong & \hat{\tau}_{\mathrm{\iptw}}(\lambda_{1}^{*},\lambda_{2}^{*})\\
 &  & -E\left\{ \frac{\partial\hat{\tau}_{\iptw}(\lambda_{1}^{*},\lambda_{2}^{*})}{\partial(\lambda_{1},\lambda_{2})^{\T}}\right\} E\left\{ \frac{\partial\hat{Q}(\lambda_{1}^{*},\lambda_{2}^{*})}{\partial(\lambda_{1},\lambda_{2})^{\T}}\right\} ^{-1}\hat{Q}(\lambda_{1}^{*},\lambda_{2}^{*})\\
 & = & \hat{\tau}_{\mathrm{\iptw}}(\lambda_{1}^{*},\lambda_{2}^{*})-B_{1}^{\T}\hat{Q}_{1}(\lambda_{1}^{*},\lambda_{2}^{*})-B_{2}^{\T}\hat{Q}_{2}(\lambda_{1}^{*},\lambda_{2}^{*})\\
 & = & \frac{1}{n}\sum_{i=1}^{m}\sum_{j=1}^{n_{i}}\left\{ \alpha_{ij}(\lambda_{1}^{*},\lambda_{2}^{*})A_{ij}(Y_{ij}-B_{1}^{\T}X_{ij})+B_{1}^{\T}X_{ij}\right\} \\
 &  & -\frac{1}{n}\sum_{i=1}^{m}\sum_{j=1}^{n_{i}}\left\{ \alpha_{ij}(\lambda_{1}^{*},\lambda_{2}^{*})(1-A_{ij})(Y_{ij}-B_{2}^{\T}X_{ij})+B_{2}^{\T}X_{ij}\right\} \\
 & = & \frac{1}{n}\sum_{i=1}^{m}\sum_{j=1}^{n_{i}}\tau_{ij},
\end{eqnarray*}
where
\begin{multline*}
\tau_{ij}=\left\{ \alpha_{ij}(\lambda_{1}^{*},\lambda_{2}^{*})A_{ij}(Y_{ij}-B_{1}^{\T}X_{ij})+B_{1}^{\T}X_{ij}\right\} \\
-\left\{ \alpha_{ij}(\lambda_{1}^{*},\lambda_{2}^{*})(1-A_{ij})(Y_{ij}-B_{2}^{\T}X_{ij})+B_{2}^{\T}X_{ij}\right\} .
\end{multline*}
Therefore, $\var(\hat{\tau}_{\iptw})=\var(n^{-1}\sum_{i=1}^{m}\sum_{j=1}^{n_{i}}\tau_{ij})$,
denoted as $V_{1}$. 

To establish the asymptotic normality of $\hat{\tau}_{\iptw}$, we
use the central limit theory for dependent variables \citep{hoeffding1948central,serfling1968contributions}.
Let $\var(\tau_{ij})=\sigma_{\tau}^{2}$ and $\mathrm{cov}(\tau_{ij},\tau_{ik})=\nu_{\tau}$
for $j\neq k$. Arrange the $\tau_{ij}$'s in a $n$-length sequence
$\{\tau_{11},\ldots,\tau_{1n_{1}},\tau_{21},\ldots,\tau_{mn_{m}}\}$.
To simplify the notation, let the $k$th random variable in this sequence
be denoted by $\tau_{k}$, for $k=1,\ldots,n$. We now consider such
sequences $\{\tau_{k}:k=1,\ldots,n\}$ are indexed by $n$. By Condition
\ref{RegCondition} (iv), the absolute central moments $E|\tau_{k}-E(\tau_{k})|^{3}$
are bounded uniformly in $k$. Moreover, by the assumption of $\sup_{1\leq i\leq m}n_{i}=O(n^{1/2})$,
we then have $\var(\sum_{k=a+1}^{a+n}\tau_{k})\sim nA^{2}$, uniformly
in $a$, as $n\rightarrow\infty,$ where $A^{2}$ is a positive constant.
Following \citet{serfling1968contributions}, these are typical criterion
for verifying the Lindeberg condition \citep{loeve1960probability},
and therefore $V_{1}^{-1/2}(\hat{\tau}_{\iptw}-\tau)\rightarrow\mathcal{N}(0,1)$,
in distribution, as $n\rightarrow\infty$.

\bibliographystyle{dcu}
\bibliography{ci}

\end{document}